\documentclass[letterpaper]{IEEEtran}
\usepackage{amsmath,amsfonts}
\usepackage{mathrsfs}
\usepackage{algorithmic}
\usepackage{array}
\usepackage{textcomp}
\usepackage{siunitx}
\usepackage{stfloats}
\usepackage{url}
\usepackage{subcaption}
\usepackage{verbatim}
\usepackage{graphicx}
\usepackage[hidelinks]{hyperref}
\def\BibTeX{{\rm B\kern-.05em{\sc i\kern-.025em b}\kern-.08em
    T\kern-.1667em\lower.7ex\hbox{E}\kern-.125emX}}
\usepackage{balance}

\begin{document}
\title{Channel Modeling and Experimental Validation of Odor-Based Molecular Communication Systems}

% End-to-End Experimental Validation of Odor-Based MC

\newcommand{\orcidiconFeb}{\href{https://orcid.org/0009-0008-2632-1140}{\includegraphics[scale=0.1]{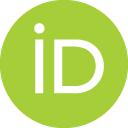}}}

\newcommand{\orcidiconAbk}{\href{https://orcid.org/0009-0006-1187-7782}{\includegraphics[scale=0.1]{figures/orcidID128.png}}}

\newcommand{\orcidiconOba}{\href{https://orcid.org/0000-0003-2523-3858}{\includegraphics[scale=0.1]{figures/orcidID128.png}}}

\author{Ahmet B. Kilic\orcidiconAbk,~\IEEEmembership{ Student Member,~IEEE}, 
        Fatih E. Bilgen\orcidiconFeb,~\IEEEmembership{Graduate Student Member,~IEEE},
        and Ozgur B. Akan\orcidiconOba,~\IEEEmembership{Fellow,~IEEE}     
        \thanks{Ahmet B. Kilic, Ozgur B. Akan are with the Center for neXt-generation Communications (CXC), Department of Electrical and Electronics Engineering, Koç University, Istanbul, Turkey (e-mail: \{ahmetkilic20, akan\}@ku.edu.tr).}
        \thanks{Fatih E. Bilgen, Ozgur B. Akan are with the Internet of Everything (IoE) Group, Electrical Engineering Division, Department of Engineering, University of Cambridge, Cambridge, CB3 0FA, UK (email: \{feb49, oba21\}@cam.ac.uk).}
	    \thanks{This work was supported in part by the AXA Research Fund (AXA Chair for Internet of Everything at Ko\c{c} University).}
}

\maketitle

\begin{abstract}
Odor-based Molecular Communication (OMC) employs odor molecules to convey information, contributing to the realization of the Internet of Everything (IoE) vision. Despite this, the practical deployment of OMC systems is currently limited by the lack of comprehensive channel models that accurately characterize particle propagation in diverse environments. While existing literature explores various aspects of molecular transport, a holistic approach that integrates theoretical modeling with experimental validation for bounded channels remains underdeveloped. In this paper, we address this gap by proposing mathematical frameworks for both bounded and unbounded OMC channels. To verify the accuracy of the proposed models, we develop a novel experimental testbed and conduct an extensive performance analysis. Our results demonstrate a strong correlation between the theoretical derivations and experimental data, providing a robust foundation for the design and analysis of future end-to-end OMC systems.
\end{abstract}


\section{Introduction}
\IEEEPARstart{M}{olecular} communication (MC) refers to the transfer of information through particles such as molecules \cite{akanmcsurvey}. When odor molecules are employed to carry messages, the process is specifically termed odor-based molecular communication (OMC). 

% Inspired by biological phenomena, the development of OMC systems holds significant promise. These systems offer transformative potential across various sectors, including biomedical engineering, industrial and consumer goods, environmental monitoring, telecommunications, and future Internet of Everything (IoE) applications \cite{aktas2025odorbased}.

Inspired by biological phenomena, OMC systems hold promise for IoE applications spanning sensing, monitoring, and short-range communication. However, realizing this potential fundamentally depends on the accurate modeling and analysis of end-to-end OMC systems. In this context, the development of realistic channel models is of central importance.

Deriving these models necessitates a combination of theoretical analysis, simulation, and experimental validation. Although numerous studies have investigated various combinations of these methods \cite{farsad2013tabletop, channelmodelexp}, a comprehensive and integral approach to modeling bounded channels remains absent from the literature. To address this gap, this study presents a mathematical model for both bounded and unbounded channel characteristics. We subsequently validate and verify this framework using a novel experimental testbed.

The remainder of this paper is structured as follows: Section \ref{sec: section 2} presents the mathematical models for both bounded and unbounded channels. Section \ref{sec: section 3} describes the developed experimental setup and methodology. In Section \ref{sec: section 4}, we analyze the experimental results and evaluate their correlation with the theoretical models derived in Section \ref{sec: section 2}. Finally, Section \ref{sec: section 5} provides concluding remarks.

\section{System Model}
\label{sec: section 2}

In this section, we establish the analytical framework for an end-to-end odor-based molecular communication (OMC) system. The proposed model comprises a point-source transmitter, an air-based propagation channel, and a metal-oxide (MOX) sensor acting as the receiver. We formulate the particle propagation in continuous time and space to derive the channel characteristics for unbounded and bounded environments.

\subsection{Transmitter}
%We model the transmitter as an ideal point source emitting odor molecules instantaneously at time $t=0$. The transmitter is located at the origin of the Cartesian coordinate system, i.e., at $(x,y,z)=(0,0,0)$. 
The transmitter is modeled as an instantaneous point source at the origin. Under this convention, the source term in the governing advection--diffusion equation is expressed as
\begin{equation}
\label{eq:diracem}
S(x,y,z,t)
=
M\,\delta(x)\delta(y)\delta(z)\delta(t),
\end{equation}
where $M$ denotes the amount of released molecules.

\begin{comment}
The transmitter releases a finite amount of odor molecules (OMs) into the channel using a spray-based mechanism. The emission is modeled as a point source in space, while its temporal characteristics depend on the modeling abstraction.

For analytical channel characterization, each transmission is approximated as an instantaneous point release at $t=0$, yielding the channel impulse response. This idealized emission is expressed as
\begin{equation}
\label{eq:diracem}
S(x,y,z,t)
=
M\,\delta(x)\delta(y)\delta(z)\delta(t),
\end{equation}
where $S(x, y, z, t)$ denotes the source term for the advection-diffusion equation and $M$ denotes the total number of released molecules.
\end{comment}

\subsection{Channel Impulse Response}
The propagation of odor molecules is modeled as an advection and diffusion process in a three-dimensional medium
\begin{equation}
\frac{\partial C}{\partial t}
=
K \nabla^2 C
-
\nabla \cdot (\mathbf{u} C)
+
S,
\label{eq:pde}
\end{equation}
where $C$ is the concentration function in space-time, $K$ is the effective (turbulent) diffusion coefficient, $\mathbf{u}$ is the flow velocity vector, and $S$ is the source term.

\begin{comment}
Odor propagation is governed by advection and diffusion in a three-dimensional medium. The concentration field $C(\mathbf{x},t)$ satisfies the advection--diffusion equation
\begin{equation}
\frac{\partial C}{\partial t}
=
K \nabla^2 C
-
\nabla \cdot (\mathbf{u} C)
+
S,
\label{eq:pde}
\end{equation}
where $K$ denotes the effective (turbulent) diffusion coefficient, $\mathbf{u}$ is the flow velocity vector, and $S$ represents the source term.
\end{comment}

%In the considered mathematical model, the flow field and boundary conditions are time-invariant, while the source term varies in time. The channel therefore operates under quasi-steady conditions and is linear and time-invariant (LTI) with respect to the emitted odor waveform, enabling impulse-response-based modeling and temporal superposition.

\subsubsection{Unbounded Medium}
In the unbounded case, the medium is assumed infinite in the horizontal directions and semi-infinite in the vertical direction, with a perfectly reflecting ground plane at $z=-h$. The flow velocity is constant and aligned with the $x$-axis, $\mathbf{u}=(u,0,0)$. Under the assumption of instantaneous release, we adopt the closed-form concentration response derived in \cite{Bige2016}, expressed as
\begin{equation}
\begin{aligned}
C_{\text{unb}}(x,y,z,t)
&=
\frac{M}{8(\pi r)^{3/2}}
\exp\!\left(-\frac{(x-ut)^2+y^2}{4r}\right)
\\&\left[
\exp\!\left(-\frac{z^2}{4r}\right)
+
\exp\!\left(-\frac{(z+2h)^2}{4r}\right)
\right],
\end{aligned}
\label{eq:unbounded}
\end{equation}
where $h$ denotes the source height from the ground and
\begin{equation}
r(x)
=
\frac{1}{u}
\int_0^x
K(\xi)\, d\xi.
\end{equation}
For constant turbulent diffusivity, this reduces to $r = Kx/u$. 

\subsubsection{Bounded Medium}
The transmitter is located at the geometric center of the cross section, i.e., at $(y,z)=(0,0)$, consistent with the source definition in \eqref{eq:diracem}. To model propagation within a channel of square cross-section (edge length $2l$, bounded at $y, z = \pm l$), we solve the governing PDE under the source term $S(x,y,z,t)$ defined in \eqref{eq:diracem}. Assuming a constant wind $u$ along the $x$-axis and neglecting longitudinal diffusion due to advection dominance, the PDE can be expressed as
\begin{equation}
    \frac{\partial C}{\partial t} + u\frac{\partial C}{\partial x} - K(x) \left( \frac{\partial^2 C}{\partial y^2} + \frac{\partial^2 C}{\partial z^2} \right) = S(x,y,z,t).
\label{eq:pde_bounded}
\end{equation}

Following \cite{Bige2016}, the bounded-channel impulse response is obtained by solving the transverse diffusion equation under Neumann boundary conditions. The resulting concentration field is expressed as
\begin{equation}
C_{bnd}(x,y,z,t) = \frac{M}{u} a(r,y) b(r,z) \delta(t-r),
\label{eq:bounded_impulse}
\end{equation}
where
\begin{equation}
a(r,y) = \frac{1}{2l} + \frac{1}{l}\sum_{n=1}^\infty e^{-(n\pi/l)^2 r} \cos\left(\frac{n\pi y}{l}\right),
\end{equation}
and $b(r,z)$ is defined analogously. By symmetry, $b(r,z)$ is identical to $a(r,y)$ with $z$ replacing $y$. 

\subsection{Channel Pulse Response}
The closed-form concentration functions derived thus far rely on the assumption
of an instantaneous release to simplify the solution of the governing PDE.
However, the experimental transmitter emits odor molecules as a pulse with a
finite duration, denoted by $T_p$. Since the channel operates under quasi-steady
conditions and is modeled as a linear time-invariant (LTI) system, the response
to this finite-duration release is obtained through the temporal superposition
of the channel impulse response. The LTI assumption holds in expectation under
statistically stationary airflow conditions, as observed in our experimental
study. Accordingly, for the unbounded medium, the concentration field is given by
\begin{equation}
C_{\text{unb}}^{(T_p)}(x,y,z,t)
=
\frac{1}{T_p}
\int_{0}^{\min(t,T_p)}
C_{\text{unb}}(x,y,z,t-\tau)\, d\tau,
\label{eq:conv_unb}
\end{equation}
where $C_{\text{unb}}(x,y,z,t)$ denotes the unbounded-channel impulse response in \eqref{eq:unbounded}. Similarly, for the bounded medium, the concentration field is given by
\begin{equation}
C_{\text{bnd}}^{(T_p)}(x,y,z,t)
=
\frac{1}{T_p}
\int_{0}^{\min(t,T_p)}
C_{bnd}(x,y,z,t-\tau)\, d\tau,
\label{eq:conv_bnd}
\end{equation}
where $C_{bnd}(x,y,z,t)$ is the bounded-channel impulse response given in \eqref{eq:bounded_impulse}.

\subsection{Modeling of Reception}

The receiver maps the ambient concentration at the receiver location,
$C_{\text{ambient}}(t)$, into a voltage signal through the static sensitivity
and dynamic kinetics of a metal-oxide (MOX) sensor. The steady-state sensing
resistance follows a power-law relationship \cite{channelmodelexp}
\begin{equation}
R_{\text{static}}(t) = R_0 \cdot 10^b \cdot [C_{\text{ambient}}(t)]^m,
\end{equation}
where $R_0$ is the reference resistance and $m$ and $b$ are datasheet-derived
sensitivity parameters \cite{PololuMQ3Datasheet2026}. The corresponding
equilibrium output voltage is obtained via a voltage divider
\begin{equation}
V_{\text{static}}(t) = V_c \frac{R_L}{R_L + R_{\text{static}}(t)},
\end{equation}
where $V_c$ is the constant circuit voltage  and $R_L$ is the load resistance. Due to adsorption--desorption kinetics, the measured voltage exhibits temporal
hysteresis and is modeled as a first-order system
\begin{equation}
\frac{dV(t)}{dt} = \frac{1}{\tau(t)} \left( V_{\text{static}}(t) - V(t) \right),
\end{equation}
with a piecewise time constant
\begin{equation}
\tau(t)=
\begin{cases}
\tau_{\text{rise}}, & V_{\text{static}}(t) > V(t),\\
\tau_{\text{decay}}, & V_{\text{static}}(t) \le V(t).
\end{cases}
\end{equation}

Measurement noise is modeled as a signal-dependent Gaussian process
\begin{equation}
V_N(t) = V(t) + \mathcal{N}(0, \kappa^2 V(t)^2),
\end{equation}
 where $\kappa$ is the noise scaling coefficient.

\section{Experimental Setup}
\label{sec: section 3}
\begin{figure*}[t]
    \centering
    \begin{subfigure}{0.25\linewidth}
        \centering
        \includegraphics[height=5cm, width=\linewidth, keepaspectratio]{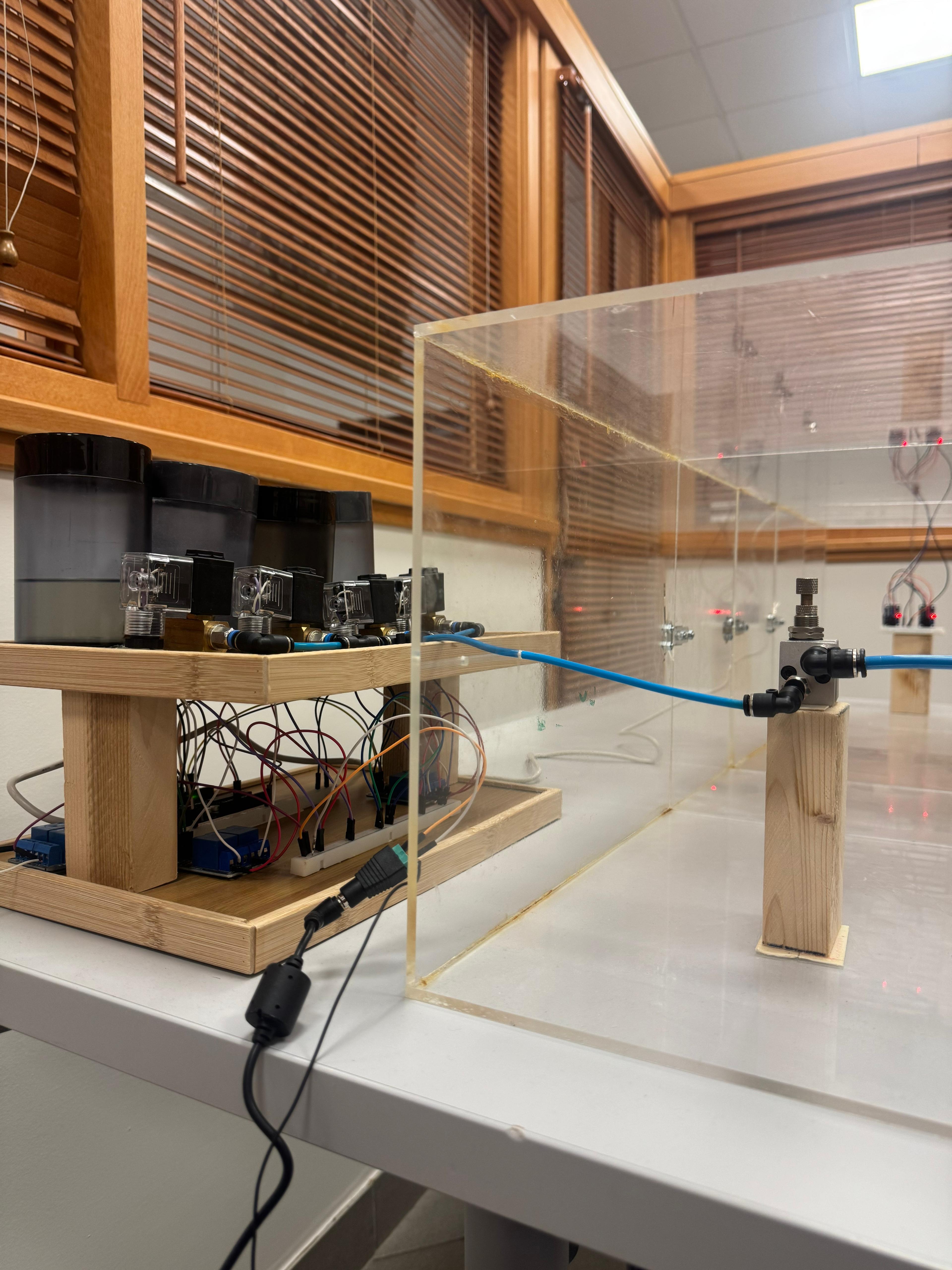}
        \caption{}
        \label{fig:exptrans}
    \end{subfigure}
    \begin{subfigure}{0.25\linewidth}
        \centering
        \includegraphics[height=5cm, width=\linewidth, keepaspectratio]{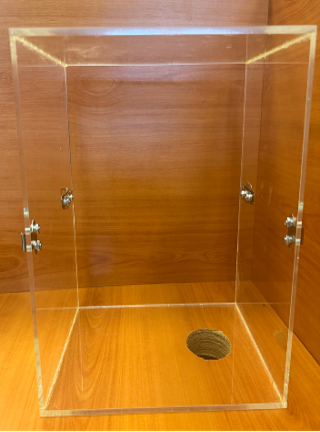}
        \caption{}
        \label{fig:expchan}
    \end{subfigure} 
    \begin{subfigure}{0.25\linewidth}
        \centering
        \includegraphics[height=5cm, width=\linewidth, keepaspectratio]{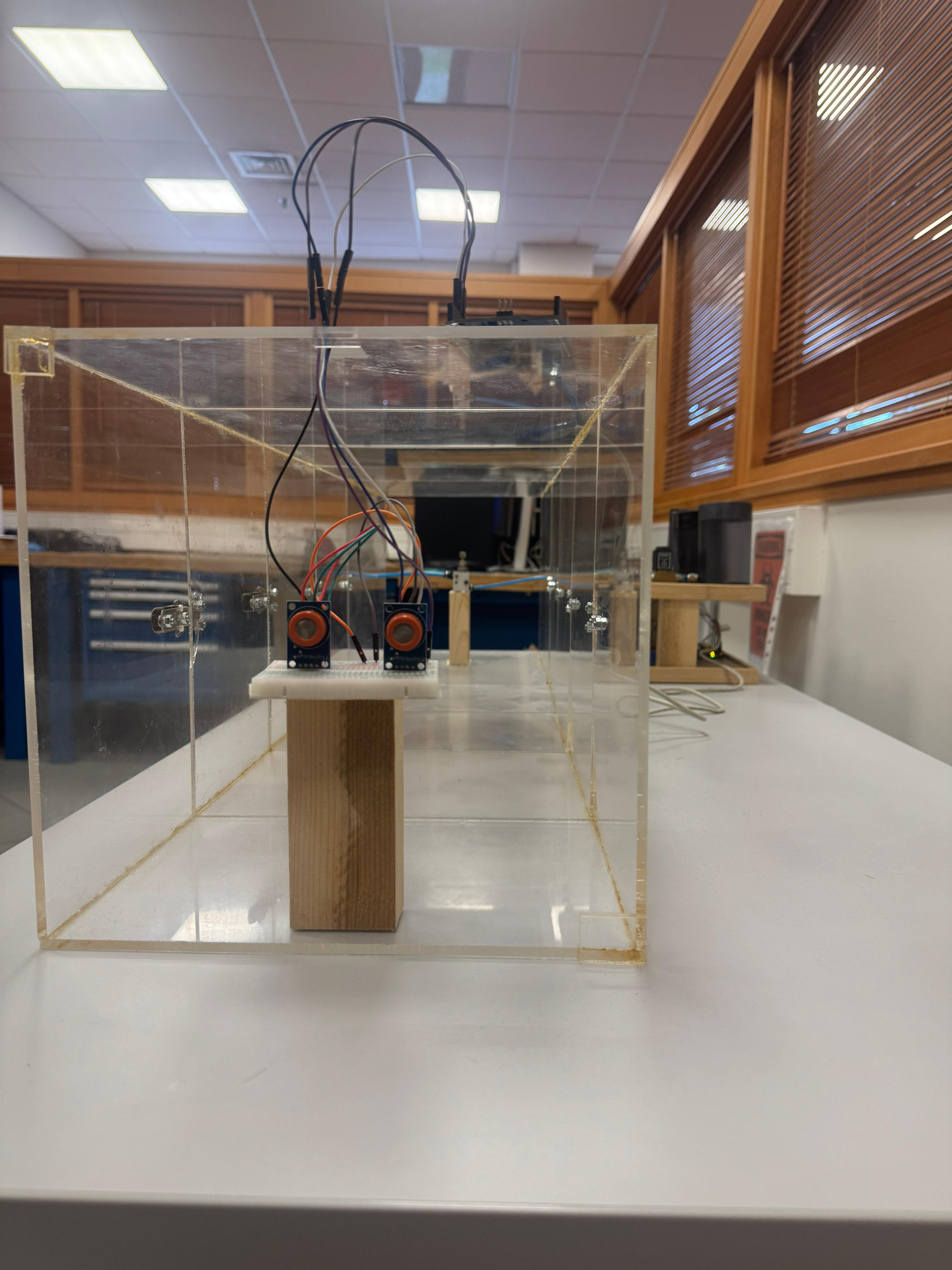}
        \caption{}
        \label{fig:exprec}
    \end{subfigure}
    \caption{Experimental setup for the propagation channel (a) transmitter, (b) channel, and (c) receiver.}
    \label{fig:expsetup}
\end{figure*}

This section outlines the hardware implementation of the experimental testbed used to verify the proposed mathematical models. The experimental setup consists of three primary components: a custom-built transmitter, a reconfigurable propagation channel, and a receiver module featuring a Metal-Oxide (MOX) sensor. The components of experimental setup are shown in Fig.~\ref{fig:expsetup}.

\subsection{Transmitter}
The transmitter is a custom-designed odor delivery system capable of encoding digital information into chemical signals. It comprises four distinct reservoirs, each capable of holding a separate liquid solution, as shown in Fig.~\ref{fig:exptrans}. These reservoirs are connected to a common output nozzle via solenoid valves controlled by an Arduino Uno microcontroller. To facilitate atomization, a constant airflow of 2 bar is provided by an air compressor exclusively during the odor release intervals. This pressure forces the liquid through the spray nozzle, creating a fine mist. The resulting mean flow velocity at the channel inlet was measured experimentally and fixed at $\vec{u} = (5.0,0,0)$ \si{\meter\per\second}. Through the microcontroller, any solenoid valve can be opened or closed for any duration, allowing for flexible signal generation.

\subsection{Channel}
The propagation medium is configurable to emulate both unbounded and bounded environments, allowing for a comprehensive analysis of diverse channel characteristics.

\subsubsection{Unbounded Medium}
To simulate an unbounded environment, the transmitter releases odor molecules into an open laboratory space. While there are no lateral or vertical physical boundaries immediately constraining the plume, the experimental apparatus is positioned on a laboratory table. This surface acts as a reflective ground plane ($z=-h$), which is explicitly accounted for in the theoretical model.

\subsubsection{Bounded Medium}
For bounded channel experiments, a modular plexiglass tunnel is utilized. The channel is constructed from $25 \times 25$ cm square cross-section segments, as shown in Fig.~\ref{fig:expchan}. These segments can be interlocked to form a tunnel up to 2 meters in length. This setup creates a confined environment where reflections from the walls ($y = \pm l, z = \pm l$) significantly impact the concentration profile, mirroring the theoretical model derived for bounded propagation.

\subsection{Receiver}
The receiver unit employs two MQ-3 MOX gas sensors to detect the concentration of odor molecules. These sensors are directly interfaced with an Arduino Uno microcontroller for data acquisition, as shown in Fig.~\ref{fig:exprec}.  This dual-sensor configuration is implemented to enhance system reliability by reducing hardware-dependent failures and mitigating measurement noise through signal averaging.

\section{Analysis and Results}
\label{sec: section 4}

This section validates the proposed channel and receiver models using experimental data. System parameters are first established through calibration, after which the model accuracy is evaluated for single-pulse transmissions, multi-pulse sequences, and noise behavior.

\subsection{Parameter Selection and Estimation}

Accurate representation of the experimental testbed requires careful identification of both physical channel parameters and receiver characteristics.

\subsubsection{Physical Parameters}

The physical parameters are determined based on the airflow regime and the geometry of the experimental setup. With a mean flow velocity of $u = 5$~m/s, the channel operates in a high Reynolds number regime, where transport is dominated by turbulent mixing rather than molecular diffusion. Accordingly, an effective diffusivity of $K = 0.05$~\si{\meter\squared\per\second} is employed \cite{MohdJuffry2023}.

%The released mass $M$ is estimated from the physical characteristics of the odor emission mechanism.  Ethanol is discharged through a circular orifice of diameter $d = 0.5$~cm at the base of a cylindrical reservoir. Assuming gravity-driven flow with an effective hydraulic head of $h = 9$~cm, the volumetric flow rate is approximated using Torricelli’s law as
% \[
% Q = A \sqrt{2 g h} \approx 2.60 \times 10^{-5}\ \text{m}^3/\text{s}.
% \]
% For a minimum valve actuation time of $T_{\text{pulse}} = 1$~s, the released volume is $V_r \approx 26.0$~cm$^3$. Accounting for a 10\% transport loss and an ethanol concentration of 80\% by volume, the effective released molar quantity is
% \[
% M = \frac{\rho_{\text{eth}} \cdot 0.8 \cdot 0.9 \cdot V_r}{M_{w,\text{eth}}} \approx 0.32\ \text{mol},
% \]
% where $\rho_{\text{eth}} = 0.789$~g/cm$^3$ and $M_{w,\text{eth}} = 46.07$~g/mol.

Each transmission releases approximately $26.0$~cm$^3$ of ethanol. Using the ethanol density and molar mass \cite{PubChemEthanol2026}, this corresponds to
a released amount of approximately $0.32$~mol.

\subsubsection{Receiver Calibration}

Receiver parameters are obtained from calibration measurements and datasheet specifications. The reference resistance $R_0$ is estimated by normalizing the clean-air sensor resistance $R_{s,\text{air}}$ using the manufacturer-provided low-concentration intercept $\Gamma \approx 60$ \cite{PololuMQ3Datasheet2026}, yielding $R_0 \approx R_{s,\text{air}}/60$. The static sensor response is modeled by the power-law relation $R_s \propto 10^b C^m$. The sensitivity slope is fixed at $m = -1.03$, while the intercept $b$ is chosen to match the datasheet reference point at a concentration of $0.4$~mg/L.

The rise time constant is consistently small, with $\tau_{r} \approx 0.05$~s for the unbounded case and $\tau_{r} \approx 0.23$~s for the bounded case, indicating rapid adsorption dynamics. In contrast, the decay time constant is noticeably larger in the unbounded experiments ($\tau_{d} \approx 45.0$~s) than in the bounded case ($\tau_{d} \approx 30.0$~s). Since unbounded trials were conducted sequentially following the bounded experiments without intermediate sensor regeneration, this increase is attributed to sensor fatigue. Residual ethanol accumulation on the sensing layer likely slowed desorption kinetics during later measurements.

The receiver coordinates $(x_r, y_r, z_r)$ are selected to align with the source in $y$- and $z-$ axes in order to maximize signal capture. All system parameters used in the analysis are summarized in Table~\ref{tab:parameters}.

% \begin{table}[h!] \centering \caption{System Parameters} \begin{tabular}{|l|c|l|c|} \hline Parameter & Symbol & Value & Ref. \\ \hline Released Mass & $M$ & 0.32 \si{\mol} & \cite{PubChemEthanol2026} \\ \hline Effective Diffusivity & $K$ & $0.05$ \si{\meter\squared\per\second} & \cite{MohdJuffry2023} \\ \hline Flow Velocity & $\vec{u}$ & $(5.0, 0, 0)$ \si{\meter\per\second} & - \\ \hline Source Height & $h$ & $0.125$ \si{\meter} & - 
% \\ \hline Receiver Coordinate & $(x_r,y_r,z_r)$ & ($1.10,0,0.125$)\si{\meter} & - 
% \\ \hline Channel Half-Width & $l$ & $0.125$ \si{\meter} & - \\ \hline Reference Resistance & $R_0$ & 302.82 \si{\ohm} & - \\ \hline Sensitivity Slope & $m$ & $-1.03$ & \cite{channelmodelexp} \\ \hline Sensitivity Intercept & $b$ & $0.40$ & - \\ \hline Circuit Voltage & $V_S$ & $5.0$ \si{\volt} & - \\ \hline Load Resistance & $R_L$ & $20$ \si{\kilo\ohm} & \cite{PololuMQ3Datasheet2026} \\ \hline Rise Time (Unbounded) & $\tau_{r,unb}$ & $0.05$ \si{\second} & - \\ \hline Rise Time (Bounded) & $\tau_{r,bnd}$ & $0.23$ \si{\second} & - \\ \hline Decay Time (Unbounded) & $\tau_{d,unb}$ & $45.0$ \si{\second} & - \\ \hline Decay Time (Bounded) & $\tau_{d,bnd}$ & $30.0$ \si{\second} & - \\ \hline Noise Scaling Factor & $\kappa$ & 0.01 & - \\ \hline \end{tabular} \label{tab:parameters} \end{table}

\begin{table*}[t]
\centering
\caption{System Parameters}
\label{tab:parameters}
\setlength{\tabcolsep}{6pt} % Adjust padding for cleaner look
\renewcommand{\arraystretch}{1.1}
\begin{tabular}{|l|c|l||l|c|l||l|c|l|}
\hline
\textbf{Parameter} & \textbf{Sym.} & \textbf{Value} & \textbf{Parameter} & \textbf{Sym.} & \textbf{Value} & \textbf{Parameter} & \textbf{Sym.} & \textbf{Value} \\ \hline
Released Mass & $M$ & 0.32 mol & Ref. Resistance & $R_0$ & 302.8 $\Omega$ & Rise Time (Unb) & $\tau_{rise}$ & 0.05 s \\
Diffusivity & $K$ & 0.05 m$^2$/s & Sens. Slope & $m$ & $-1.03$ & Rise Time (Bnd) & $\tau_{rise}$ & 0.23 s \\
Flow Velocity & $\vec{u}$ & $(5.0, 0, 0)$ m/s & Sens. Intercept & $b$ & $0.40$ & Decay Time (Unb) & $\tau_{decay}$ & 45.0 s \\
Source Height & $h$ & 0.125 m & Circuit Voltage & $V_c$ & 5.0 V & Decay Time (Bnd) & $\tau_{decay}$ & 30.0 s \\
Rx Coordinate & $\vec{r}_{rx}$ & $(1.10, 0, 0)$ m & Load Resistance & $R_L$ & 20 k$\Omega$ & Noise Factor & $\kappa$ & 0.01 \\
Half-Width & $l$ & 0.125 m & Pulse Duration & $T_p$ & 1 \si{\second} & - & - & - \\
\hline
\end{tabular}
\end{table*}

\subsection{One-Shot Response}

\begin{figure*}[t]
    \centering
    \begin{subfigure}{0.4\linewidth}        \includegraphics[height=4cm, width=\linewidth, keepaspectratio]{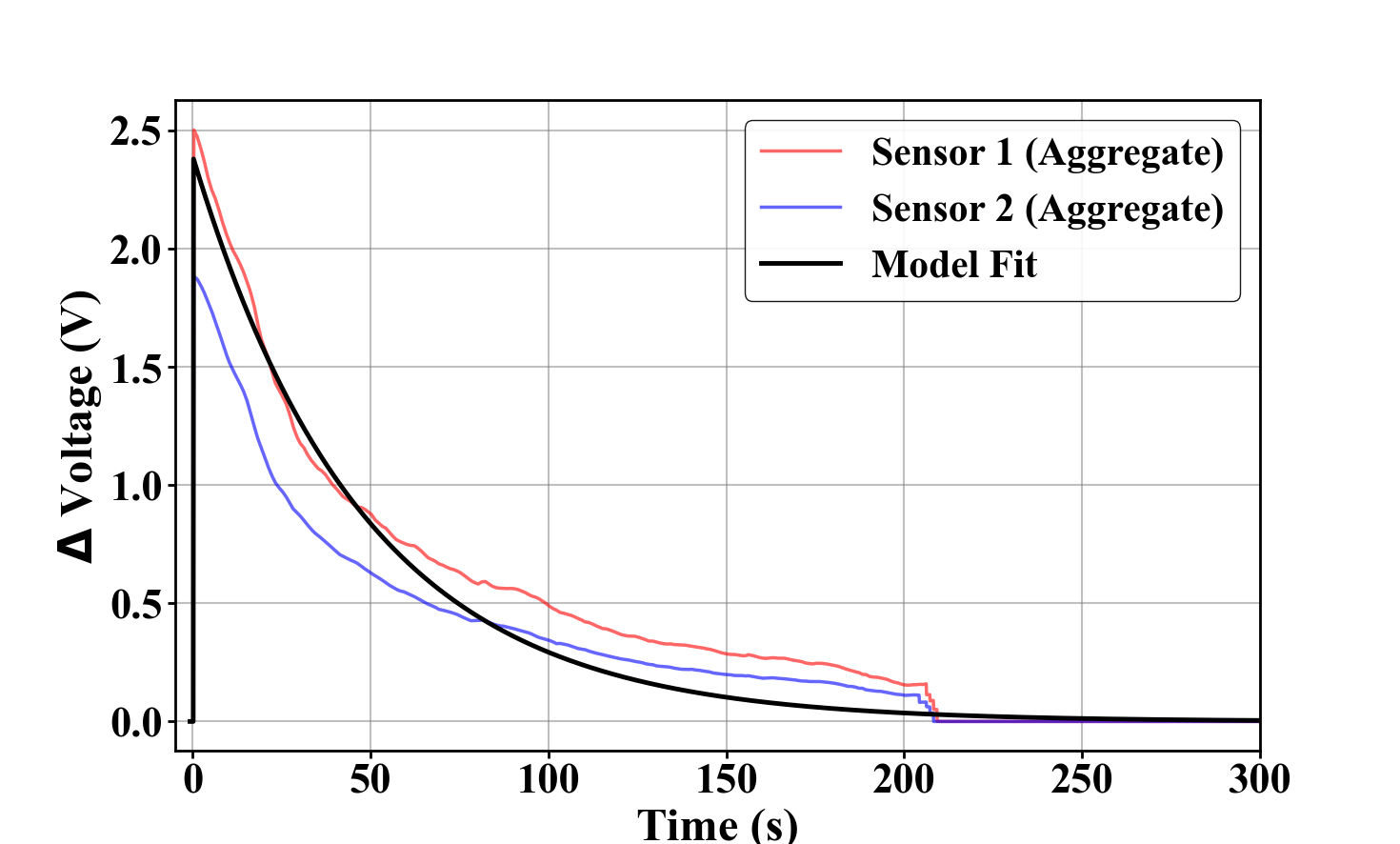}
        \caption{}
        \label{fig:boundedoneshot}
    \end{subfigure}
    \begin{subfigure}{0.4\linewidth}
        \includegraphics[height=4cm, width=\linewidth, keepaspectratio]{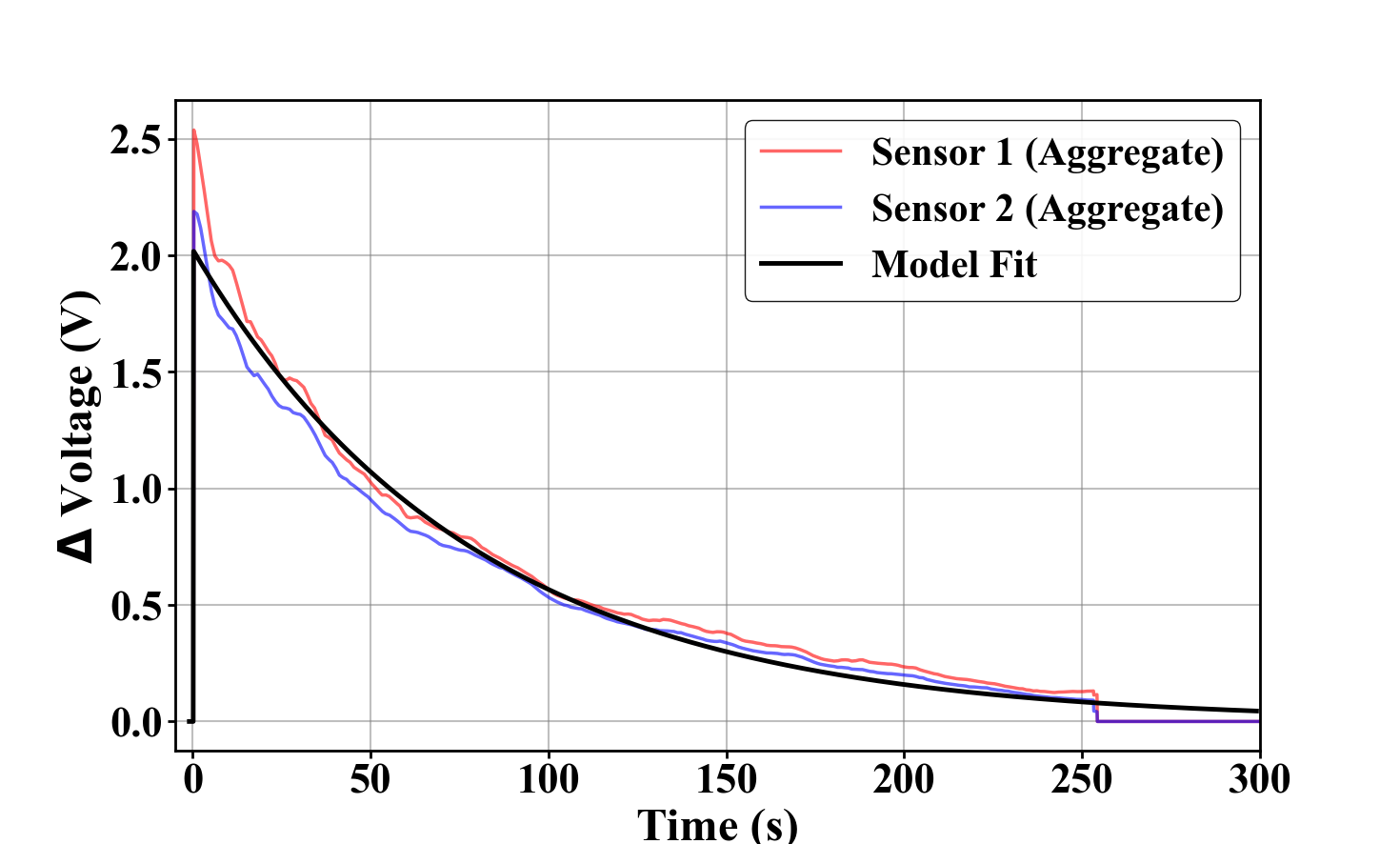}
        \caption{}
        \label{fig:unboundedoneshot}
    \end{subfigure} 
    \caption{Experimental validation of the single-shot response for (a) bounded and (b) unbounded propagation environments.}
    \label{fig:one_shot_results}
\end{figure*}

The channel pulse response is evaluated by transmitting a single $1.0$~s ethanol pulse in both bounded and unbounded configurations. Averaged experimental voltage traces are compared with the corresponding theoretical predictions in Fig.~\ref{fig:one_shot_results}.

In the bounded channel (Fig.~\ref{fig:boundedoneshot}), the response exhibits a pronounced peak followed by a relatively rapid return toward the baseline. This behavior suggests that the confining boundaries of the tunnel, combined with the directed airflow, maintain a coherent plume structure that is efficiently cleared from the sensor vicinity once the main body of the puff passes. The proposed bounded-channel model accurately captures this dynamic, achieving a Pearson correlation of $r = 0.9842$ and a normalized root-mean-square error (NRMSE) of $5.40\%$. The peak amplitude error remains below $8.56\%$, indicating appropriate calibration of the released mass $M$ and longitudinal diffusivity $K$.

In contrast, the unbounded response (Fig.~\ref{fig:unboundedoneshot}) exhibits a lower peak amplitude and a significantly slower decay tail. This prolonged recovery is consistent with unconfined turbulent dispersion in open air; the absence of lateral boundaries results in lower advection gradients and allows diluted molecular clouds to linger around the receiver for extended durations. The Gaussian puff model provides an excellent match for this lingering effect, yielding a correlation of $r = 0.9922$ and a lower NRMSE of $2.73\%$. The larger peak estimation error ($14.64\%$) is attributed to the inherent variability of open-air turbulence, where stochastic eddy fluctuations can more easily disperse the high-concentration core of the plume compared to the more stabilized flow within a bounded duct.

\subsection{Multi-Shot Response and ISI}
\begin{figure*}[t]
    \centering
    % Subfigure 1
    \begin{subfigure}{0.495\linewidth}
        \centering
        % height=18cm forces the image to be tall. 
        % keepaspectratio=false allows it to stretch (distorting the image).
        % If you use the new Python plots, remove 'height' and let them scale naturally.
        \includegraphics[height=5.5cm, keepaspectratio]{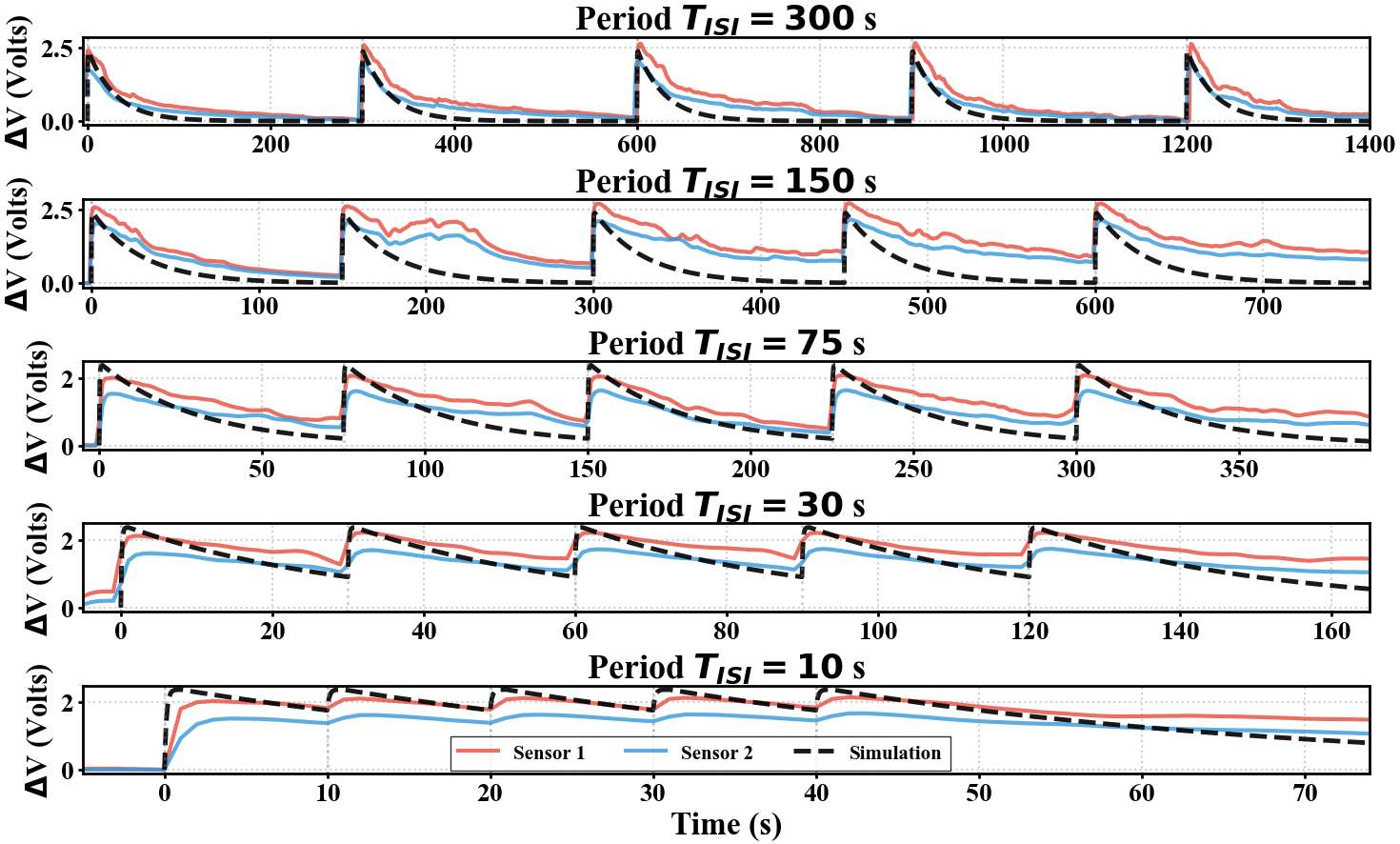}
        \caption{}
        \label{fig:boundedmultishot}
    \end{subfigure}
    \hfill % Adds space between the two subfigures
    % Subfigure 2
    \begin{subfigure}{0.495\linewidth}
        \centering
        \includegraphics[height=5.5cm, keepaspectratio]{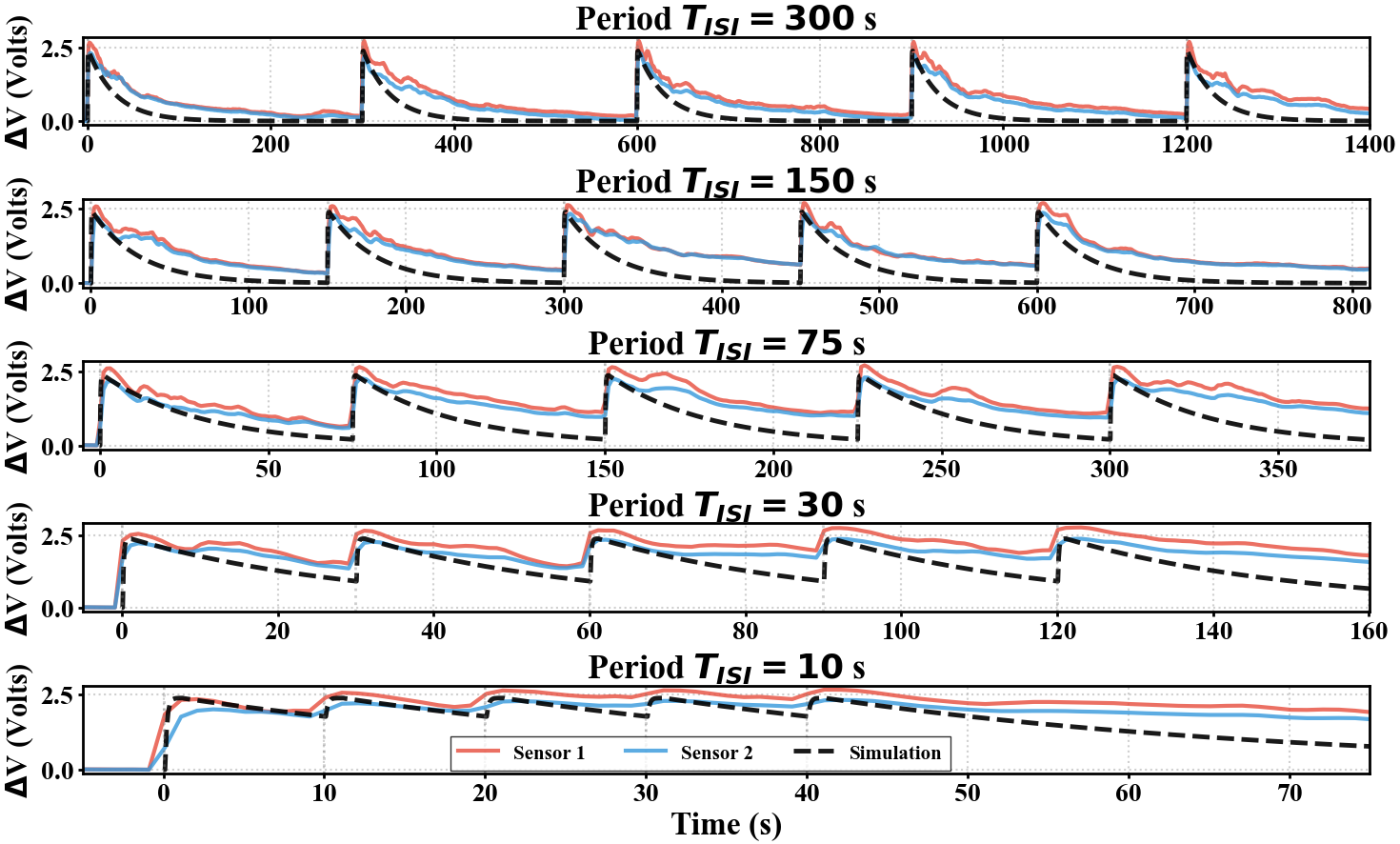}
        \caption{}
        \label{fig:unboundedmultishot}
    \end{subfigure} 
    \caption{Experimental validation of the multi-shot response for (a) bounded and (b) unbounded propagation environments. }
    \label{fig:multi_shot_results}
\end{figure*}

To evaluate inter-symbol interference (ISI) and the receiver’s dynamic behavior under continuous transmission, five-pulse sequences were transmitted with symbol periods
$T_{sym} \in \{300, 150, 75, 30, 10\}$\,s. Experimental results for bounded and unbounded configurations are shown in Fig.~\ref{fig:multi_shot_results}.

In the bounded channel (Fig.~\ref{fig:boundedmultishot}), decreasing the symbol
period $T_{sym}$ leads to a pronounced transition from isolated pulse responses
to strong inter-symbol interference (ISI). At long symbol periods
($T_{sym}=300$\,s), the sensor output largely returns to baseline between
transmissions, and model agreement is moderate due to slow environmental drift
and weak temporal coupling between symbols. As $T_{sym}$ decreases, residual
odor accumulation becomes dominant, and the response exhibits clear cumulative
buildup. This behavior arises from boundary-induced plume confinement combined
with the sensor’s slow desorption dynamics
($\tau_{\text{decay}} \approx 30$\,s). Importantly, model accuracy improves
monotonically with decreasing $T_{sym}$, with correlation increasing from
$r \approx 0.55$ at $T_{sym}=300$\,s to above $r=0.97$ at $T_{sym}=10$\,s. This
indicates that predominantly deterministic ISI accumulation at high transmission
rates is well captured by the convolution-based channel model, whereas low-rate
operation is more affected by uncontrolled environmental variability.

In contrast, the unbounded channel (Fig.~\ref{fig:unboundedmultishot}) exhibits
substantially reduced ISI across all symbol periods. Free transverse diffusion
facilitates effective plume dissipation between successive transmissions,
limiting baseline drift even at short $T_{sym}$. As a result, model agreement
remains consistently high over the entire range of symbol periods
($r > 0.9$ in most cases), with only weak dependence on transmission rate. The
absence of sustained odor accumulation in the unbounded configuration indicates
that channel confinement, rather than sensor kinetics, is the dominant factor
limiting high-rate transmission in bounded environments.

\subsection{Noise Characterization}

\begin{figure*}[t]
    \centering
    \begin{subfigure}{0.485\linewidth}
        \includegraphics[height=4cm, keepaspectratio]{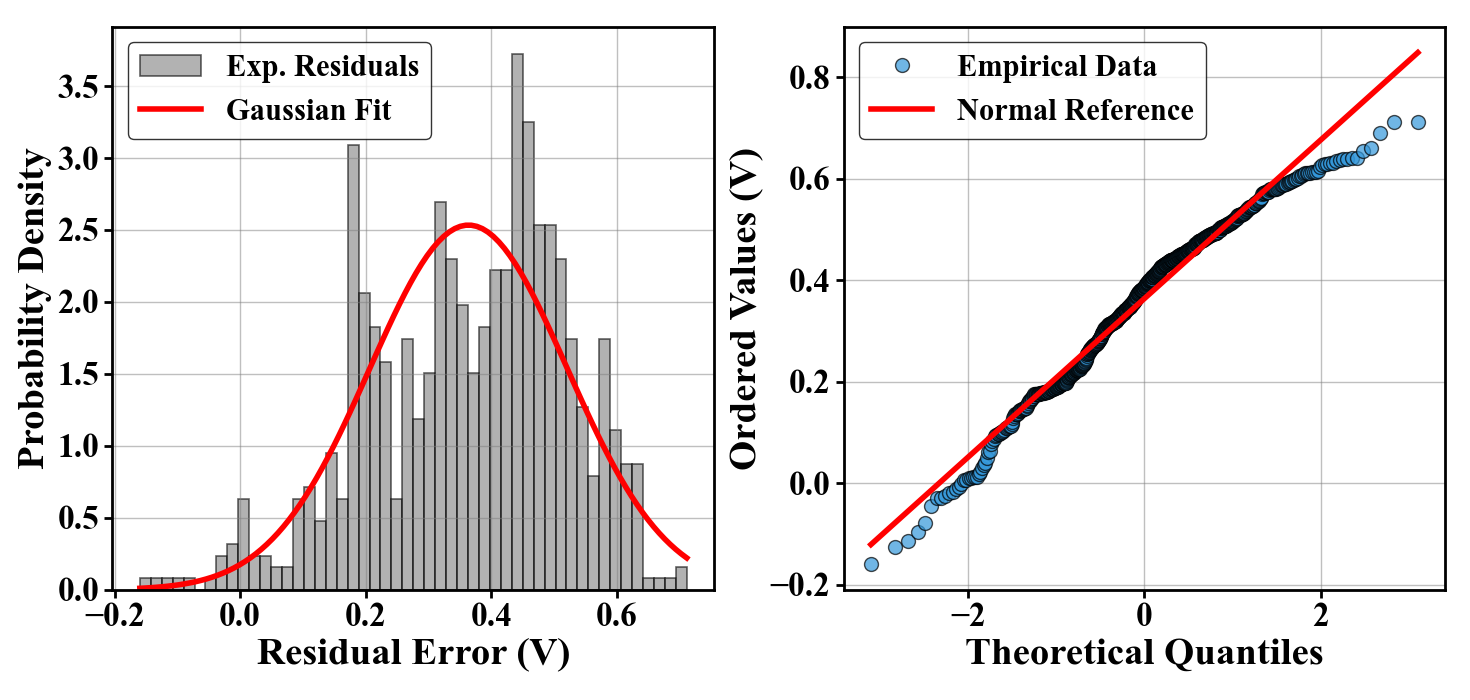}
        \caption{}
        \label{fig:boundednoise_char}
    \end{subfigure}
    \begin{subfigure}{0.485\linewidth}
        \includegraphics[height=4cm, keepaspectratio]{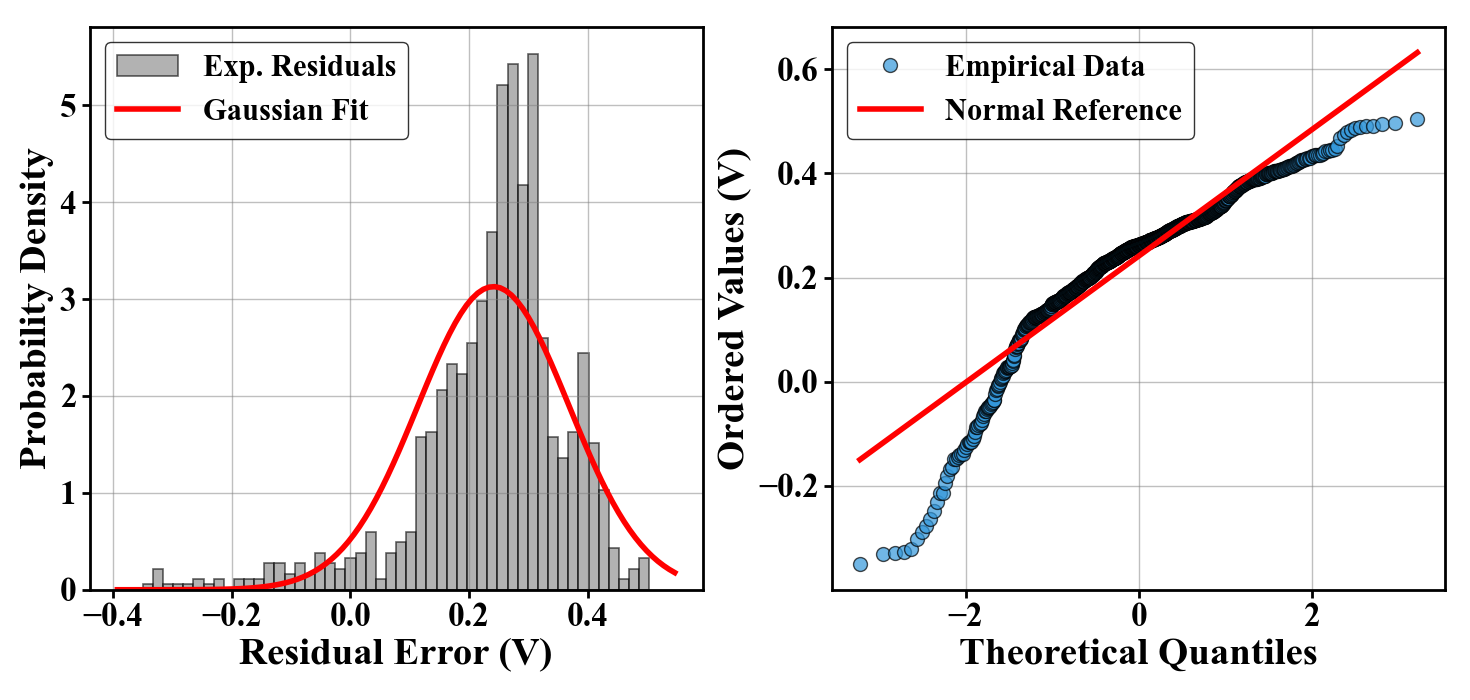}
        \caption{}
        \label{fig:unboundednoise_char}
    \end{subfigure} 
    \caption{Noise characterization of (a) bounded and (b) unbounded propagation environments.}
    \label{fig:noise_char}
\end{figure*}

To assess the validity of the receiver noise model, the residual error between measurements and model predictions was analyzed, defined as
$\epsilon(t) = V_{exp}(t) - V_{model}(t)$ where $V_{model}(t)$ corresponds to the noise-free voltage $V(t)$ predicted by the receiver model. The residual distributions for bounded and unbounded configurations are shown in Fig.~\ref{fig:noise_char}.

For the unbounded channel (Fig.~\ref{fig:unboundednoise_char}), the residual histogram is approximately symmetric and centered at zero, consistent with a Gaussian distribution. The corresponding Q--Q plot exhibits strong linearity over the central quantiles ($\pm 1\sigma$), indicating good agreement with the normal model. A deviation in the lower tail (quantiles $< -1$) is observed, which is indicative of turbulent intermittency. While the Gaussian model predicts a smooth mean concentration, the instantaneous turbulent field contains pockets of clean air that momentarily reduce the local concentration, causing the measured signal to fall below the model prediction and resulting in negative residuals.

The bounded channel residuals (Fig.~\ref{fig:boundednoise_char}) also exhibit a dominant Gaussian core, though with slightly increased irregularity. This behavior is consistent with the more complex mixing dynamics induced by reflective boundaries, which can sustain localized concentration variations. The Q--Q plot shows minor deviations in the lower tail, likely caused by the sensor’s asymmetric recovery dynamics during plume clearance.

Overall, the close agreement between the empirical residuals and the Gaussian reference across the bulk of the data supports the use of an additive Gaussian noise model for subsequent analysis and performance evaluation.

\section{Conclusion}
In this paper, we have presented a comprehensive framework for the modeling and experimental validation of end-to-end Odor-based Molecular Communication (OMC) systems. We derived analytical characterizations for molecular propagation in both bounded and unbounded environments, explicitly integrating the kinetic dynamics of metal-oxide (MOX) sensors. Through the development of a custom experimental testbed, we verified that our theoretical models exhibit a strong correlation with physical measurements across single-shot and multi-shot transmission scenarios. By accurately capturing channel behavior, inter-symbol interference (ISI), and measurement noise, this study provides a robust foundation for the design and practical deployment of future OMC systems.

\label{sec: section 5}

\bibliographystyle{IEEEtran}
\bibliography{references.bib}

% \begin{IEEEbiography}[{\includegraphics[width=1in,height=1.25in,clip,keepaspectratio]{fig1.png}}]{IEEE Publications Technology Team}
% In this paragraph you can place your educational, professional background and research and other interests.\end{IEEEbiography}

\end{document}